# Demonstration of imaginary-mass particles by optical simulation in non-Hermitian systems

**Authors:** Wange Song[1,2], Shenglun Gao[1,2], Hanmeng Li[1,2], Chen Chen[1,2], Shengjie Wu[1,2], Shining Zhu[1,2] & Tao Li[1,2,*]

## Affiliations

[1]National Laboratory of Solid State Microstructures, Key Laboratory of Intelligent Optical Sensing and Manipulation, Jiangsu Key Laboratory of Artificial Functional Materials, College of Engineering and Applied Sciences, Nanjing University, Nanjing, 210093, China.

[2]Collaborative Innovation Center of Advanced Microstructures, Nanjing, 210093, China

*Correspondence to: taoli@nju.edu.cn (T.L.)

## ABSTRACT

Unifying quantum mechanics and special relativity, the Dirac equation describes the behaviour of relativistic quantum particles, including imaginary-mass particles with faster-than-light speeds (*e.g.*, tachyon). However, experimental searches for such particles remain negative so far. Fortunately, there are possibilities to investigate the relativistic quantum effects by quantum simulations using a controllable system. However, these simulations are mostly explored in closed physical systems that hinders the demonstration of imaginary-mass particles, whereas the general open framework would promise more possibilities in quantum simulations. Here, we reveal that the dynamic behaviors of imaginary-mass particles can be mapped to the transfer of wave packet in a non-Hermitian silicon optical lattice. We experimentally demonstrate a super-divergent dynamics of tachyon that travels faster than massless particles (*e.g.*, photon). Besides, our simulation suggests another kind of imaginary-mass particles that have rarely been explored, which show non-divergent behaviors with imaginary energy. It is the first experimental quantum simulation being able to observe the imaginary-mass particles that have never been experimentally verified. Our work provides a fully controllable and extensible platform to investigate relativistic quantum phenomena at a chip-scale level, which would inspire more insightful explorations in quantum effects with non-Hermiticity.

Keywords: quantum simulation, imaginary-mass particles, tachyon, relativistic quantum particles, non-Hermiticity, parity-time symmetry

## INTRODUCTION

As one of the breakthroughs of theoretical physics, Dirac equation describes massive particle-antiparticle pairs (*i.e.*, antimatter) and massless fermions (*e.g.*, neutrinos and anti-neutrinos) [1-3]. Importantly, the Dirac theory also suggests potentially particles with imaginary mass that travel faster than light, *i.e.* tachyon [4]. The topic of this faster-than-light particle once has received considerable attention from theoretical physicists in the 1960s [5-9]. Although the existence of tachyon-like behaviors may not conflict with the causality despite the superluminal behavior [8-10], experimental searches for tachyon remain negative so far [11-14]. Fortunately, the emulation of relativistic quantum phenomena by artificial optical materials has become accessible in recent years. With their help, various relativistic phenomena, which the direct measurement is far beyond present experiment techniques, can be observed in tabletop experiments, such as Zitterbewegung [15,16], Klein tunneling [17,18], and massless Dirac dynamics [19,20]. Cooperated with the advances of developing relativistic quantum theory, these simulations arouse the passion of exploring new striking quantum phenomena and greatly advance the study of the relativistic quantum effects.

On the other hand, advances in non-Hermitian and parity-time symmetry (PT) systems have attracted great attention with interesting effects and functionalities enabled by exceptional points (EPs) [21-25]. Broadening the range of quantum mechanics, the non-Hermiticity and EPs in open systems bring richer physics in optical simulations of nonrelativistic phenomena [26-30]. It is of interest to extend the quantum simulation of the relativistic phenomena, that is, the Dirac equation to a more general open system. For example, massless Dirac particles exhibit nondispersive conical diffractions governed by the linear dispersion around the singularity in the band structure, *i.e.*, Dirac point (DP) [31-33]. When adding non-Hermitian components, DPs are transformed into the non-Hermitian version with new striking topological signatures, such as exceptional ring [34], bulk Fermi arc [35], and topological half-charges [36]. Therefore, the non-Hermiticity introduces a new dimension in engineering the photonic band structures [37-39], and opens the door to many possibilities in optical quantum simulations related with relativistic quantum effects.

In this work, we propose and realize the dynamics of imaginary-mass particles (IMP), including tachyon, and imaginary-energy particles (IEP), in non-Hermitian waveguide lattices on a photonic chip. Through a direct analogy, we found that the Hamiltonian of optical fields in the non-Hermitian waveguide lattice, under certain conditions, can be mapped to the Dirac Hamiltonian for relativistic particles. Due to the high controllability over the on-site energy and the coupling among waveguides of the non-Hermitian lattice, the dynamics of relativistic particles can be mimicked using sinusoidally curved silicon waveguides covered with chrome (Cr) stripe [19,30,40]. Comparing to conical diffraction evolutions of massless particles, the dynamic behaviors of tachyon in non-Hermitian system exhibits super-divergent evolutions that travels faster than massless particles. This phenomenon deviates from the intuition that the non-Hermiticity always causes diffusion of light. We also successfully observe another kind of IMP with imaginary energy (IEP) that have rarely been explored. Experimental results reveal the non-divergent evolutions of IEP with imaginary speed. To the best of our knowledge, this is the first experiment to simulate the evolutions of imaginary-mass particles. Our work, therefore, deepens the understanding of dynamic behaviors of relativistic quantum particles and provides a flexible platform to further investigate the relativistic quantum effects with non-Hermiticity.

## RESULTS AND DISCUSSION

### Analogy for relativistic particles in 1D non-Hermitian waveguide lattice

Figure 1 schematically shows the one-dimensional (1D) optical lattice with alternating coupling coefficients of the nearest neighbors ($\kappa_1$ and $\kappa_2$) and PT-symmetric potential. The system can be described by the following tight-binding coupled-mode equations:

$$\begin{cases} (i\dfrac{\partial}{\partial z}+\beta_B)\psi_{2n}+\kappa_2\psi_{2n-1}+\kappa_1\psi_{2n+1}=0 \\ (i\dfrac{\partial}{\partial z}+\beta_A)\psi_{2n+1}+\kappa_1\psi_{2n}+\kappa_2\psi_{2n+2}=0 \end{cases}, \tag{1}$$

where $\psi_n$ denotes the optical field in the $n$th waveguide, $\beta_A$ and $\beta_B$ (where $\beta_A=\beta_0+i\gamma/2$, $\beta_B=\beta_0-i\gamma/2$, $\gamma$ is gain-loss strength) are the onsite propagation constants (*i.e.*, onsite energy) in lossy (A) and gain (B) waveguides, respectively. The Hamiltonian of the non-Hermitian waveguide lattice in momentum space writes

$$H_{WG}=\begin{pmatrix} i\gamma/2\kappa_1 & \kappa_2/\kappa_1 e^{-ikd}+1 \\ \kappa_2/\kappa_1 e^{ikd}+1 & -i\gamma/2\kappa_1 \end{pmatrix}, \tag{2}$$

where $k$ is the quasimomentum in the Brillouin zone and $d$ is the unit cell dimension. In the Hermitian system, the bandgap closes and a singular point appears with a linear relationship around the point [*i.e.*, Dirac point (DP)]. However, the non-Hermitian effects change the band structure significantly and split the Dirac point into a pair of EPs (see **Supporting Information I** for details).

Now we reveal a direct analogy between the Hamiltonian of the waveguide lattice and that of relativistic particles. According to 1D Dirac equations, the Dirac Hamiltonian can be written as

$$H_D=\sigma_1 cp+\sigma_3 mc^2, \tag{3}$$

with eigenvalues $E^2=m^2c^4+p^2c^2$, where $\sigma_i$ refers to Pauli matrices, $c$ is the speed of light, $m$ is the mass of the relativistic particle, and $p$ is the momentum. In analogy to the Dirac Hamiltonian, the Hamiltonian $H_{WG}$ in Eq. (2), around $k$=0, can be transformed into the familiar form of $H_D$ through proper transformation (see **Supporting Information II** for details)

$$H_a=\sigma_1(-\kappa_2/\kappa_1)^{1/2}kd+\sigma_3[(1+\kappa_2/\kappa_1)^2-(\gamma/2\kappa_1)^2]^{1/2}\equiv\sigma_1 c_a p_a+\sigma_3 m_a c_a^2. \tag{4}$$

Here, we set $(-\kappa_2/\kappa_1)^{1/2}\equiv c_a$, $kd\equiv p_a$, and $[(1+\kappa_2/\kappa_1)^2-(\gamma/2\kappa_1)^2]^{1/2}/(-\kappa_2/\kappa_1)\equiv m_a$, corresponding to the parameters of $c$, $p$, and $m$ in Eq. (3), respectively. We notice the term $[(1+\kappa_2/\kappa_1)^2-(\gamma/2\kappa_1)^2]^{1/2}/(-\kappa_2/\kappa_1)$ is in analogy to the mass term of Dirac Hamiltonian. In Hermitian case with $\gamma$=0, the term writes $(1+\kappa_2/\kappa_1)/(-\kappa_2/\kappa_1)$ and $\kappa_2/\kappa_1$=-1 represents the massless case [18,19]. If non-Hermitian parameter $\gamma$ is introduced, the circumstances become complex. Figure 1c and d show the 2D diagrams of $|\mathrm{Re}(m_a)|$ and $|\mathrm{Im}(m_a)|$ as functions of the coupling contrast ($\kappa_2/\kappa_1$) and normalized loss ($\gamma/\kappa_1$). Two white guiding lines show the boundary defined by massless particles, which follows $\kappa_2/\kappa_1=\pm\gamma/2\kappa_1-1$, it indicates that the massless particle can also exist in non-Hermitian systems. If $\gamma$ increases and crosses this boundary, $|\mathrm{Re}(m_a)|$ gets to zero while $|\mathrm{Im}(m_a)|$ becomes nonzero. It implies particles with pure imaginary mass, termed as imaginary-mass particles (IMP). Without loss of generality, we select (0, -1), (0, -0.8), and (0.3, -1) on the parameter plane ($\gamma/\kappa_1$, $\kappa_2/\kappa_1$), marked as the black, blue, and red dots in Fig. 1c corresponding to massless, massive, and IMP. The energy-momentum ($E_a$-$p_a$) relationship can be obtained by solving the eigenvalues of Hamiltonian in Eq. (4), which is

$$[(\beta-\beta_0)/\kappa_1]^2\equiv E_a^2=m_a^2c_a^4+p_a^2c_a^2. \tag{5}$$

The velocities of particles $v_a$ write

$$(v_a/c_a)^2=p_a^2c_a^2/(p_a^2c_a^2+E_{a0}^2). \tag{6}$$

The relationship of $E_a$-$p_a$ and $v_a$-$p_a$ of the particles for these cases are shown in Fig. 1e-h, respectively. Rewriting Eq. (5) as $E_a^2/(m_ac_a^2)^2-p_a^2/(m_ac_a)^2=1$, we find that $E_a$-$p_a$ has a hyperbolic relationship for massive particles ($m_a\neq0$), and upper and lower curves represent positive and negative energy states, respectively (blue curves in Fig. 1e). These are normal particles with the dispersion curve inside the light cone, indicating their speeds are always lower than light speed (blue curves in Fig. 1g). For massless particles, their dispersions become two straight lines (*i.e.*, the asymptotes of former hyperbolic curves) with a crossing at $p_a$=0, which resembles a Dirac point (black curves in Fig. 1e). The massless particles are rightly located at the light cone with light speed (black curves in Fig. 1g) [18,19]. As for the IMP that exists in non-Hermitian regimes. The non-Hermitian effect splits a DP to a pair of EPs and squeezes the $E_a$-$p_a$ curves out of the light cone (red curves in Fig. 1e). Therefore, their speeds are always larger than light speed (red curves in Fig. 1g), as termed as **tachyon**. Note that the tachyon with faster-than-light speed only exists in unbroken PT phases, where the $|p_a|$ needs to be larger than the critical EP value (*i.e.*, $|p_a|>|p_{aEP}|$, where $|p_{aEP}|=|[(1+\kappa_2/\kappa_1)^2-(\gamma/2\kappa_1)^2]/(-\kappa_2/\kappa_1)|^{1/2}$). However, within the region of $|p_a|<|p_{aEP}|$, the particle exhibits thorough imaginary energy (termed as imaginary-energy particles (IEP)), which has a non-zero value of Im($E_a$) corresponding to PT symmetry breaking (see Fig. 1f). Thus, there exist two types of IMP, one is tachyon in PT-symmetric phase that propagates faster than light, and the other is IEP existing in broken-PT phase, which has pure imaginary velocity (see Fig. 1h).

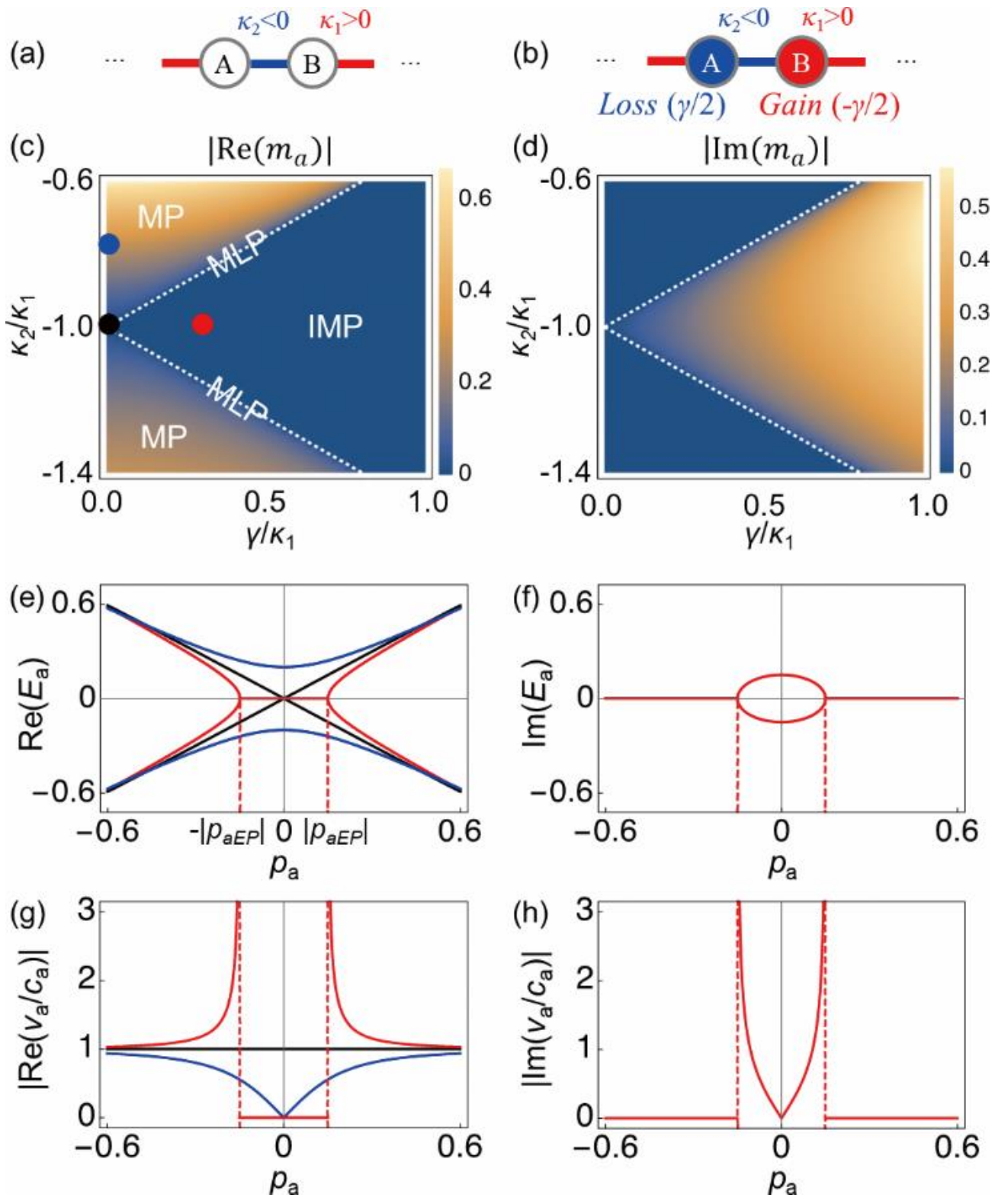

**Figure 1. Effective mass and velocities of relativistic particles by optical analogues in non-Hermitian waveguide lattice.** Illustration of 1D Hermitian (a) and non-Hermitian (b) waveguide lattice with alternating positive and negative coupling coefficients ($\kappa_1$>0 and $\kappa_2$<0). The non-Hermiticity is introduced by lossy (A sublattices) and gain (B sublattices), as indicated in blue and red circles. (c,d) $|\mathrm{Re}(m_\mathrm{a})|$ and $|\mathrm{Im}(m_\mathrm{a})|$ as functions of $\gamma/\kappa_1$ and $\kappa_2/\kappa_1$. The white dashed line represents the massless particles (MLP), distinguishing the massive particles (MP) and imaginary-mass particles (IMP). The black ($\gamma/\kappa_1$=0, $\kappa_2/\kappa_1$=-1), blue ($\gamma/\kappa_1$=0, $\kappa_2/\kappa_1$=-0.8), and red ($\gamma/\kappa_1$=0.3, $\kappa_2/\kappa_1$=-1) dots mark the MLP, MP, and IMP cases. (e,f) Real and imaginary parts of $E_\mathrm{a}$ as functions of $p_\mathrm{a}$. (g,h) Real and imaginary parts of $|v_\mathrm{a}/c_\mathrm{a}|$ as functions of $p_\mathrm{a}$. The black, blue, and red curves represent the MLP, MP, and IMP cases marked in (c). The red curves in (f) and (h) are for the IMP case, while for MLP and MP, the imaginary parts are zero and not shown here.

## Dynamics of relativistic particles

To explore the evolution of imaginary-mass particles, we consider a waveguide lattice (with 81 waveguides) for the simulation of IMP ($\kappa_1$=-$\kappa_2$=0.01 μm$^{-1}$, $\gamma$=0.003 μm$^{-1}$). Another two waveguide lattices (massless particles, $\kappa_1$=-$\kappa_2$=0.01 μm$^{-1}$, $\gamma$=0 and massive particles, $\kappa_1$=0.01 μm$^{-1}$, $\kappa_2$=-0.008 μm$^{-1}$, $\gamma$=0) are also investigated for comparison, corresponding to the cases analyzed in Fig. 1. We excite the central 21 waveguides to ensure that the distribution of the momentum remains in a narrow region centered at $\varphi_0$, where $\varphi_0$ is the phase difference of input optical fields across one unit cell, corresponding to a designed transverse $k$ vector (see **Supporting Information III** for details). Figure 2a-d show the band structure with particular excitation conditions mentioned above for massive, massless, tachyon, and IEP, respectively, and corresponding optical field dynamics are shown in Fig. 2e-h. The propagation of light in the waveguide lattice directly simulates the temporal evolution of relativistic particles, and the center of mass of the optical fields (defined as $n_\mathrm{c} = \sum_n n|\psi_n|^2 / \sum_n |\psi_n|^2$, where $n$=0,1,…,40 for the right part of the fields $n_\mathrm{cr}$ and $n$=-40,-39,…,0 for the left part $n_\mathrm{cl}$) indicates the velocity of particles. For massless particles, when an initial excitation is launched with small $\varphi_0$ into this waveguide lattice (e.g., $\varphi_0$=-0.6), the beam will split into two equally non-diffractive beams that propagate with opposite group velocities (Fig. 2f). The massless particles propagate with light speed and $n_\mathrm{cr}$=-$n_\mathrm{cl}$≈29.97 for 1600 μm according to Fig. 2i and j. For massive particles, the splitting ratio of the input fields is asymmetric for the non-zero incidence angle (see Fig. 2e) and the distance between these two parts gets smaller (Fig. 2i and j, $n_{cr}$=-$n_{cl}$≈24.55) compared with the massless case, indicating their speeds are lower than light speed. Interestingly, as the non-Hermitian modulation is introduced, the structure supports IMP. If $\varphi_0$ is small (-0.6~0.6), the distribution of the momentum is near the center of the dispersion, which corresponds to the broken-PT region. Thus IEP of IMP is excited and propagates with imaginary velocities. This in turn results in prominent non-divergent dynamics (Fig. 2h-j, $n_{cr}$=-$n_{cl}$≈14.57). However, if the input beam is titled with a $\varphi_0$ to cross the EP (e.g., $\varphi_0$=-0.6), the distribution of the momentum moves to the unbroken-PT phase region. The tachyon can be excited, which also features two equally non-diffractive beams (Fig. 2g), but propagates faster than the massless particles (Fig. 2i and j, $n_{cr}$=-$n_{cl}$≈31.22).

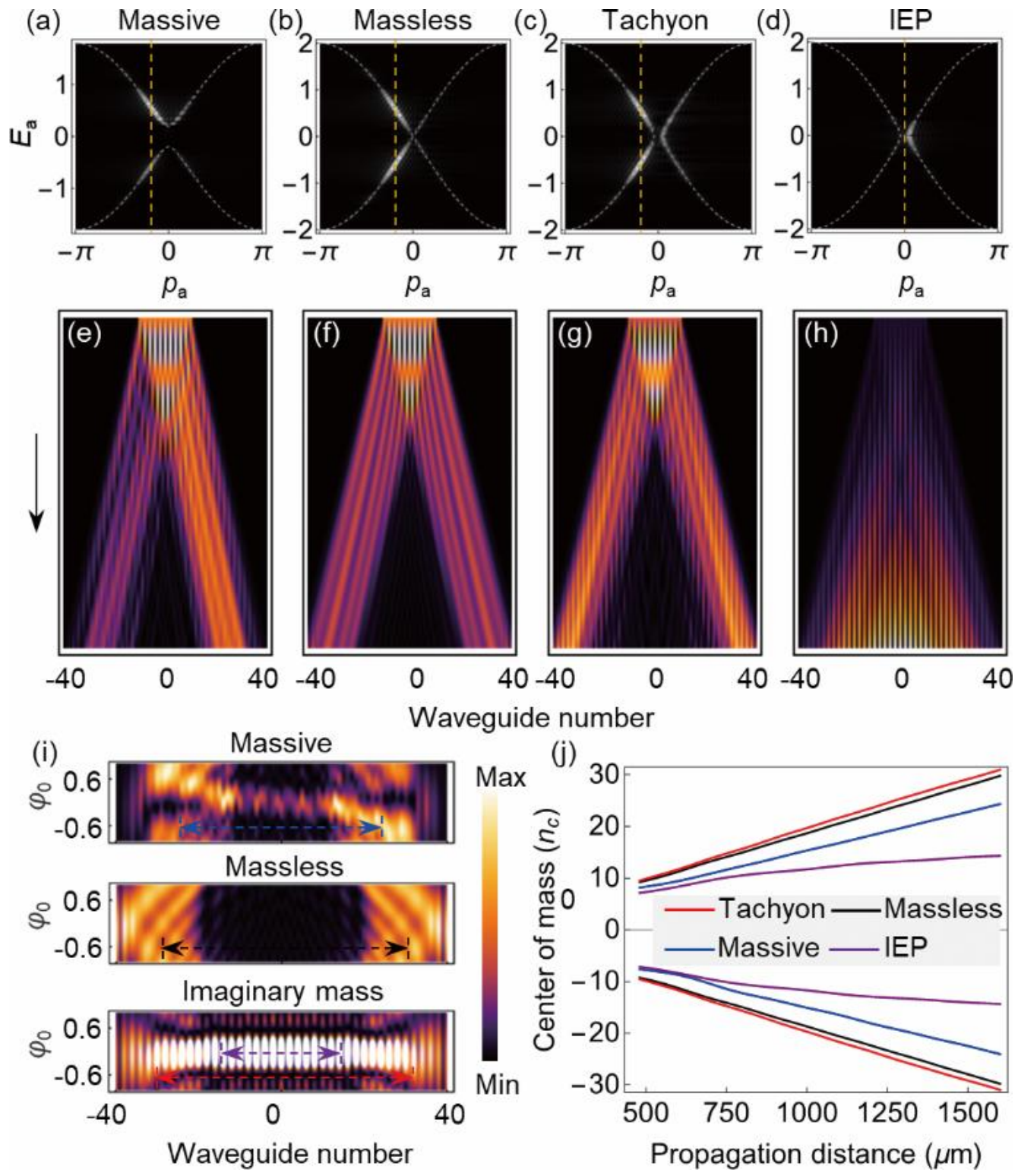

**Figure 2. Simulating dynamics of relativistic particles.** (a-d) Excited band structures for different relativistic particles, where the gray dashed lines guide the band structures, the yellow dashed lines mark the excitation position. (e-h) The corresponding evolutions of relativistic particles mimicked in non-Hermitian waveguide lattices, where the longitudinal direction (black arrow) of waveguide maps the time while the transverse maps the space. (i) Optical field distributions after evolution 1600 μm with respect to $\varphi_0$ for different relativistic particles. The blue, black, red, and purple arrows indicate the center of mass. (j) Center of mass of left and right part of (e-h).

## Numerical simulation and experimental results

To access a controllable coupling covering the positive and negative regime, we refer to a curved waveguide array, which has well proven to have the capability of coupling tuning [19,40]. Here, we carried out full-wave simulations (COMSOL MULTIPHYSICS 5.3) and experiments in curved silicon waveguides on sapphire substrate. The transverse position of the waveguide centers writes $y_0(x)=A\cos(2\pi x/P)$, where $P$ is the modulation period and $A$ is the amplitude (Fig. 3a). Figure 3b shows the simulated effective coupling coefficient as functions of $A$ and $d$ with fixed $P$=10 μm. By increasing the waveguide spacing, the effective coupling coefficient $\kappa_{eff}$ changes from positive to negative (Fig. 3c with $A$=0.68 μm). We choose alternating waveguide spacing of $d_1$=593 nm, and $d_2$=880 nm with corresponding coefficients of $\kappa_1=\kappa_{\mathrm{eff}}=0.01$ μm$^{-1}$, $\kappa_2=-\kappa_{\mathrm{eff}}=-0.01$ μm$^{-1}$ respectively, for the massless, tachyon, and imaginary energy cases. As for the massive case, we fixed $d_1$=593 nm and changed $d_2$=942 nm ($\kappa_2$=-0.008 μm$^{-1}$). Therefore, we are able to create a waveguide array with alternating positive and negative coupling by an alternation of the waveguide spacing.

Then, we introduce the non-Hermitian modulation through an array of lossy metal (*i.e.*, Chrome) nano-stripes deposited on top of every other silicon waveguide, as shown in Fig. 3a. The loss is engineered by the width of the Cr strip (see the inset of Fig. 3c). Since the full-wave simulation is quite time-consuming, we didn't perform simulations over a same large scale as the theoretical calculations. Here, we fixed the total number of waveguides to $N$=13 and 200 μm propagations. For tachyon, we designed a non-Hermitian waveguide array with 30-nm width (2-nm thickness) Cr layers, corresponding to $\gamma$=0.003 μm$^{-1}$. Due to the small broken-PT region, the IEP is difficult to excite. So we design another non-Hermitian waveguide array with larger $\gamma$ ($\gamma$=0.04 μm$^{-1}$, corresponds to 336-nm width, 2-nm thickness Cr). The field propagations of these four particles were simulated in Fig. 3h-k. Figure 3h displays the dynamics of massless particles with two non-spreading bundled rays moving away from each other, while such a splitting shrinks apparently for the massive case (see Fig. 3i). As a comparison, the field evolution of tachyon exhibits a further larger splitting angle than the massless particle (see Fig. 3j), indicating its superluminal velocity (faster-than-light speed). Finally, as $\gamma$ increases, the IEP can be observed in which a non-divergent beam dominates the dynamic pattern, and the conical diffraction effect is destructed (see Fig. 3k).

The experimental samples were fabricated by E-beam lithography and dry etching process, followed by a second-step E-beam lithography with careful alignment and lift-off process to deposit the Cr stripes (see **Methods** for details), which include the silicon waveguide array, input grating coupler, and extended output with grating couplers. We note that a trident branched waveguide was designed at the input to control the excitation $k$ vector (*e.g.,* $\varphi_0$) (see **Supporting Information IV** for details). As an example, the scanning electron microscopy (SEM) images of the fabricated tachyon structures are shown in Fig. 3d-g. In experiments, the light was input into waveguide array by focusing the laser (λ=1550 nm) via input grating coupler (see **Methods** for details). The transmitted signals can be collected from the scattered light from the output grating couplers by a near-infrared CCD camera (Xenics Xeva-1.7-320).

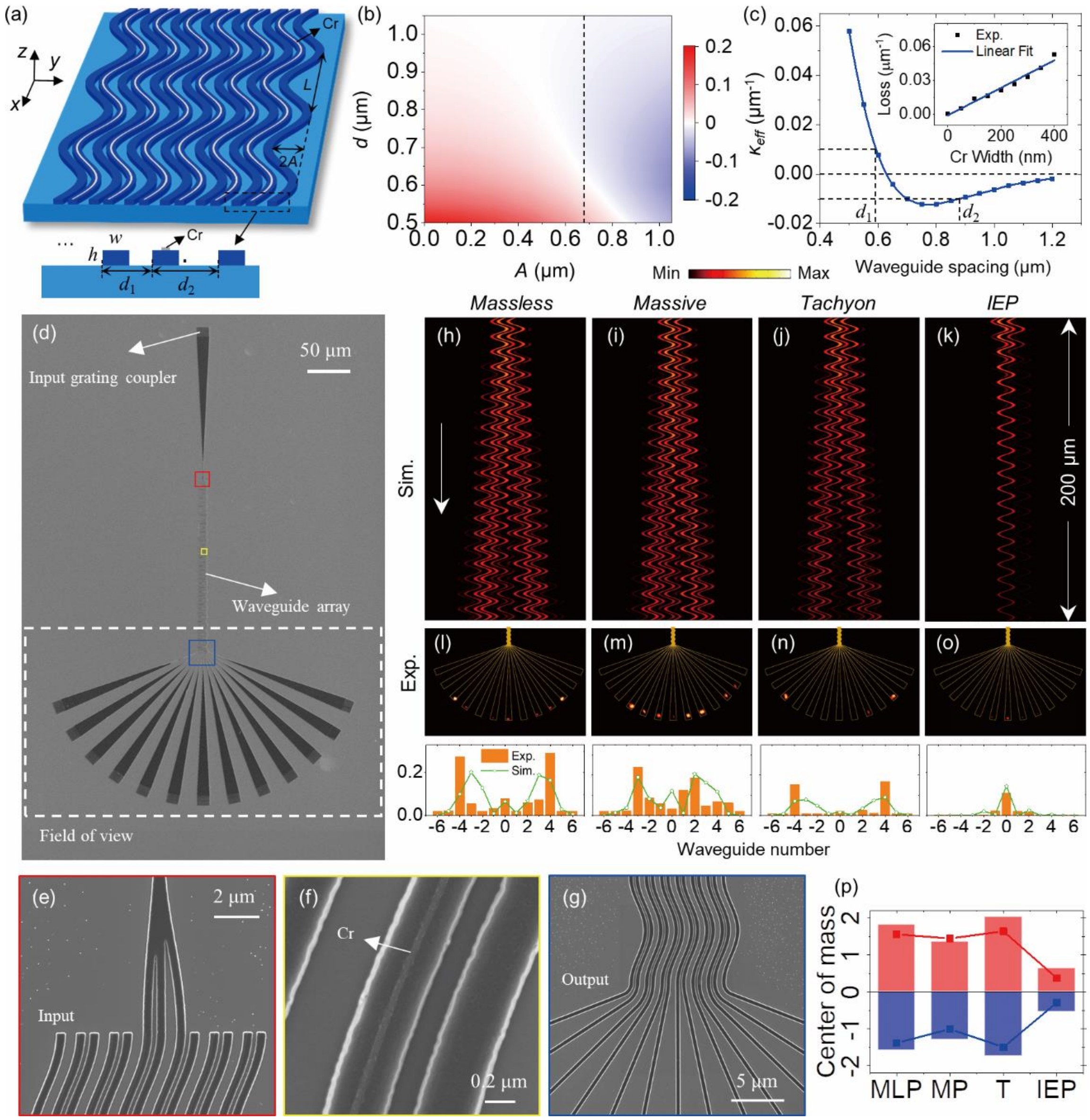

**Figure 3. Simulation and experimental results in curved silicon waveguide arrays.** (a) Schematics of the curved waveguide array consisting of the additional chrome (Cr) layer on top of every other silicon waveguide. The silicon waveguide width ($w$) and height ($h$) are optimized to support only one fundamental mode ($\beta_0$=2.1601$k_0$, $k_0$ is the free space $k$-vector) at $\lambda$=1550 nm as $w$=400 nm, $h$=220 nm. (b) Effective coupling coefficient as functions of $A$ and $d$. (c) Effective coupling coefficient as a function of waveguide spacing with $P$=10 μm and $A$=0.68 μm, where the inset depicts the loss of Cr-deposited silicon waveguide as a function of Cr width. (d-g) SEM pictures of fabricated tachyon samples. Zoomed-in images show input (e), Cr-deposited waveguides (f), and output (g) of the sample. (h-k) Simulated dynamics of different relativistic particles. (l-o) Corresponding experimentally detected output intensities (top) and normalized intensity profiles (bottom). (p) Simulated (curve) and experimental (bar) results of center of mass for relativistic particles.

Figure 3l-o display representative output scattered fields in four samples, and the corresponding extracted normalized intensities are displayed below in the bar diagrams (orange bars), which agree with the simulation results (green curves) despite the fabrication imperfections. The sample of massless particles, as the Hermitian case, exhibits two peaks in the intensity profile, corresponding to the dynamics of massless Dirac fermion (Fig. 3l). For the massive sample with unbalanced coupling, the two peaks get closer and become inconspicuous (Fig. 3m). As a comparison, the sample of tachyon with non-Hermitian modulation demonstrates two well-separated peaks in the intensity profile (Fig. 3n). This is consistent with the theoretical analysis and simulation. For the last sample of IEP with strong PT modulation, it shows an obvious localized intensity peak at the center of the waveguide array, indicating a non-divergent beam mode (Fig. 3o). We further extracted the data of center of mass from the simulated and experimental intensity distributions, as shown in Fig. 3p (curve represents the simulated data, while bar is the experimental data). It is evident that the tachyon propagates the fastest one, while the IEP has almost non-divergent evolution. Additionally, we note that the massless particles can also exist in non-Hermitian systems (see the white dashed line in Fig. 1c). The band structures of non-Hermitian massless particles also have featured linear dispersion, but their evolutions demonstrate anomalous conical diffraction in non-Hermitian systems that are completely different from the classical conical diffractions in Hermitian systems (see **Supporting Information V** for details).

So far, we have demonstrated that the dynamic behaviors of imaginary-mass particles predicted by the Dirac theory can be observed in our quantum simulation platform, but it will never be possible to achieve in real particles. Our simulation platform not only allows exceptional control of system parameters, but possesses high extendibility thanks to the flexibility and diversity of optical systems. For example, it can be generalized to study 2D versions of relativistic quantum effects by introducing synthetic dimension or using 2D waveguide lattices [41]. Besides, the emulation of interacting particles can even be possible by employing optical nonlinearity [42]. Furthermore, in the viewpoint of quantum walk, this tachyon-like behavior corresponds to a faster walker and suggests potential application in accelerating the quantum searching algorithm [43].

## CONCLUSION

In conclusion, we have demonstrated the dynamics of imaginary-mass particles predicted by the Dirac equation in curved silicon waveguide platform with a non-Hermitian configuration. Super-divergent and non-divergent evolutions are observed for superluminal tachyon and for IEP with imaginary speed, respectively, which are significantly distinct from the conventional dynamics of massless and massive particles in Hermitian systems. Since direct experimental observation of these particles is still highly intricate, it is very desirable to simulate these phenomena in a controllable systems so as to assist the advances of exploring the relativistic quantum effects. Our work reveals the strong flexibility in tailoring the coupling property as well as the dispersion engineering in waveguide array system by introducing non-Hermitian modulation, which provides an extensible platform with high level of control to mimic relativistic quantum phenomena and non-Hermitian physics on a chip.

## METHODS

### Fabrication of the photonic non-Hermitian waveguide array

The experimental samples were fabricated using the method of electron beam lithography (EBL) and dry etching process, followed by a second-step E-beam lithography with careful alignment and lift-off process to deposit the chromium (Cr) stripes. The substrate we used is 220 nm silicon deposition on 460 μm alumina substrate, which were cleaned in ultrasound bath in acetone, isopropyl alcohol (IPA) and DI water for 15 min respectively and dried under clean nitrogen flow. The alignment marks were then patterned by EBL (Elionix,

ELS-F125), deposition of chromium/gold (5 nm/40 nm) layers, and photoresist lift-off. After that, the waveguide arrays and grating nanostructures were exposed to EBL. The samples were then used to dry etch the silicon layer in a 2:5 mixture of SF6 and C4F8 plasma and the residual photoresist was stripped off by an oxygen plasma stripper. Next, the chromium structures were fabricated by an alignment E-beam lithography and lift-off process. The photoresist film was spin-coated onto the substrate. After exposing the structure, 2-nm chromium film is deposited using thermal evaporation. Then removing the photoresist by soak in N-Methyl-2-pyrrolidone (see **Supporting Information Figures S3 and S4** for details).

### Experimental characterization of the dynamics of relativistic particles.

In optical measurements, a near-infrared (NIR) optical wave at the wavelength of 1550 nm from a single frequency mode laser (FL-1550- SF) was focused at the input grating via a Mitutoyo 100X long-working distance objective (NA = 0.70, *f*=200 mm) from the front side (air). The input power was controlled with a variable neutral density (ND) filters, and the polarization of input light was tuned through a linear polarizer (LP). The scattering signals from the samples were collected using another 50X NIR objective (NA = 0.42, *f*=200 mm) in the backside (substrate), and were imaged by a NIR charge-coupled device (CCD) camera (Xenics Xeva-1.7-320) (see **Supporting Information Figure S6** for details).

## ACKNOWLEDGMENTS

We thank the micro-fabrication center of the National Laboratory of Solid State Microstructures (NLSSM) for technique support.

## FUNDING

The authors acknowledge the financial support from The National Key R&D Program of China (2017YFA0303701, 2016YFA0202103), National Natural Science Foundation of China (Nos. 91850204, 11674167). Tao Li thanks the support from Dengfeng Project B of Nanjing University.

**Conflict of interest statement. None declared.**

## AUTHOR CONTRIBUTIONS

T.L. and W.S. developed the idea. W.S. proposed the theoretical design, performed the numerical simulation; S.G. and H.L. fabricated the samples; W.S. performed the optical measurement with the help of S.W. and C.C.; T.L. organized the project. W.S. and T.L. analysed the results with support from all authors. All authors discussed the results. W.S. and T.L. wrote the manuscript with input from all authors.

## SUPPORTING INFORMATION

### I. Band structures and eigenmode properties of the non-Hermitian waveguide systems.

The non-Hermitian Hamiltonian for the model is

$$H_{WG} = \begin{pmatrix} i\gamma/2\kappa_1 & \kappa_2/\kappa_1 e^{-ikd}+1 \\ \kappa_2/\kappa_1 e^{ikd}+1 & -i\gamma/2\kappa_1 \end{pmatrix}, \tag{S1}$$

the eigenvalues can be obtained by solving Eq. (S1)

$$\begin{cases} -\dfrac{\Delta}{2\kappa_1} \\ \dfrac{\Delta}{2\kappa_1} \end{cases}, \tag{S2}$$

and corresponding eigenvectors are

$$\begin{cases} \left(\dfrac{(i\gamma-\Delta)}{2(\kappa_2 e^{ikd}+\kappa_1)},1\right) \\ \left(\dfrac{(i\gamma+\Delta)}{2(\kappa_2 e^{ikd}+\kappa_1)},1\right) \end{cases}, \tag{S3}$$

where $\Delta = \sqrt{4(\kappa_1^{\,2}+\kappa_2^{\,2})-\gamma^2+8\kappa_1\kappa_2 cos(kd)}$ .

The spectrum (Re($E$)) of $H_{WG}$ can be derived analytically in a 2D synthetic parameter space ($kd$, $\kappa_2/\kappa_1$) for Hermitian ($\gamma$=0) and non-Hermitian (e.g., $\gamma/\kappa_1$=1) cases, as shown in Figs. S1(a) and S1(b), respectively. It is evident that in the Hermitian system the bandgap closes and a singular point appears at $kd$=0, $\kappa_2/\kappa_1$=-1. The cross-sections of the band structure at $\kappa_2/\kappa_1$=-1 or $kd$=0 are shown at the bottom of Fig. S1(a) with a linear relationship around the point [*i.e.*, Dirac point (DP)]. However, in the non-Hermitian case, the singular point splits into a ring of EPs [34] [highlighted as the red loop in Fig. S1(b)]. The cross-section with Re($E$)=0 clearly presents the exceptional ring (ER) structure. For the cross-section cutting the ER, the non-Hermitian effects split the Dirac point into a pair of EPs [see cross-section with $kd$=0 or $\kappa_2/\kappa_1$=-1 in Fig. S1(b)]. Interestingly, the two points at the ER (i.e., A and B) shows the linear relationship [see cross-section with $\kappa_2/\kappa_1$=-1.5 or $kd$=0.5236 in Fig. S1(b)]. Although they have the same dispersion relationships as the DP, they are essentially EPs, because their eigenvectors are not orthogonal and become the same instead. Figures S1(c) and S1(d) show the imaginary parts for Hermitian and non-Hermitian systems.

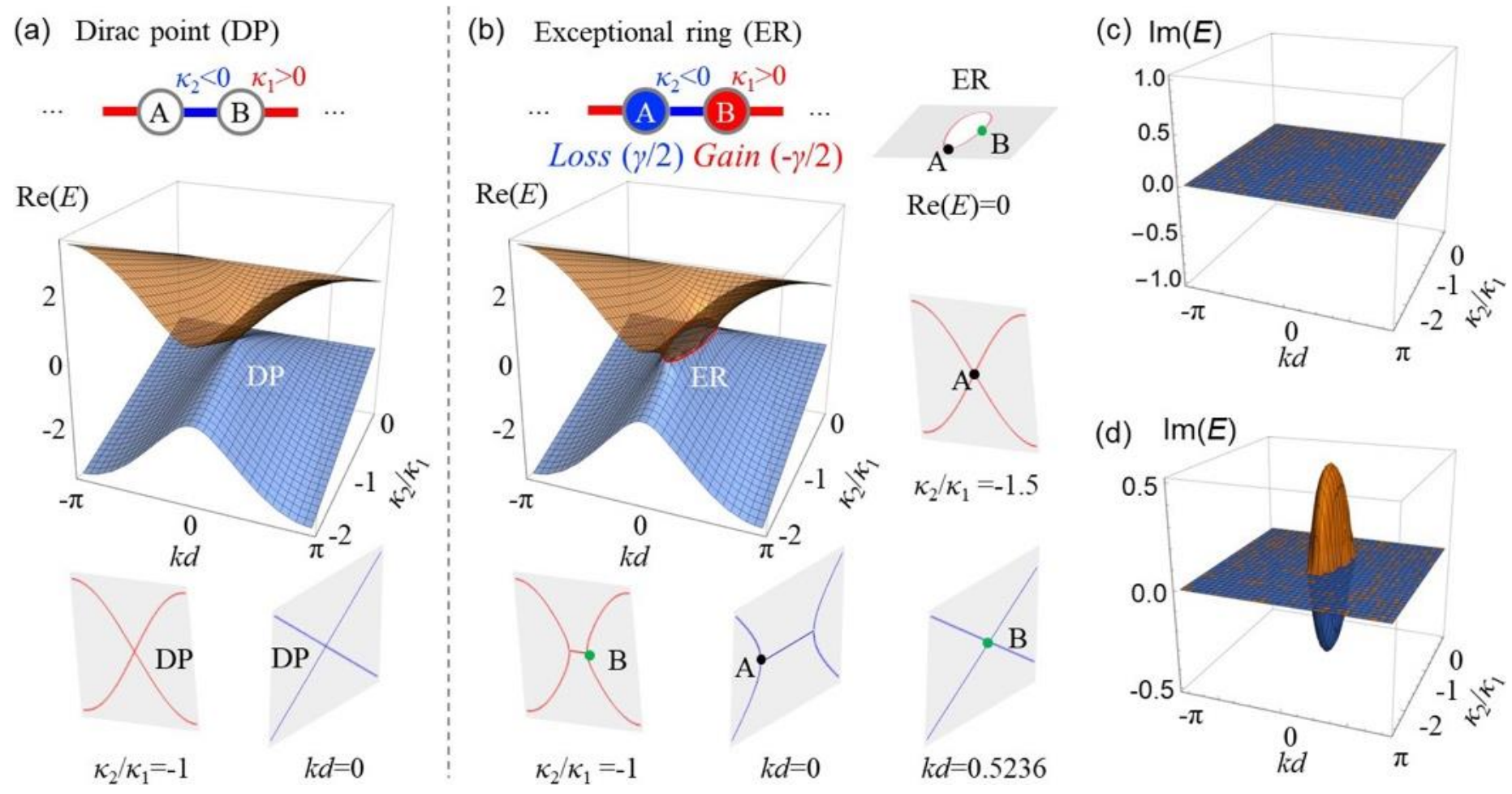

Figure S1 Exceptional ring arising from Dirac point in one-dimensional (1D) non-Hermitian waveguide lattice. Schematics of the Hermitian (a) and non-Hermitian (b) waveguide lattice. (a) Band structure of the Hermitian waveguide lattice, featuring a single Dirac point. The two insets below show the cross-section of the band structure for a given $\kappa_2/\kappa_1$=-1(red curve) and $kd$=0 (blue curve), respectively. (b) Band structure (the real part of the eigenvalues) of the non-Hermitian waveguide lattice. The Dirac point splits into a ring of exceptional points. The five insets show the sections of the non-Hermitian band structures. (c,d) The imaginary part of the eigenvalues for Hermitian (c) and non-Hermitian (d) waveguide lattices.

Figure S2 shows the band structures (a,d,g), dot product of eigenmodes of two bands ($\psi_1$. $\psi_2$) (b,e,h), and $|\psi_1$-$\psi_2|$ (c,f,i) in Hermitian ($\gamma$=0, $c_2/c_1$=-1, red curves) and non-Hermitian waveguide lattice ($\gamma$=1, $c_2/c_1$=-1, blue curves and $\gamma$=1, $c_2/c_1$=-1.5, orange curves). The band structure of Hermitian case with $\gamma$=0, $c_2/c_1$=-1 exhibits a single Dirac point with linear relationship around $k$=0 [see Fig. S2(a)]. The eigenvectors of the two bands are orthogonal across the whole band [see Fig. S2(b)]. For the non-Hermitian case with $\gamma$=1, $c_2/c_1$=-1, the Dirac point splits to paired EPs connected by a flat line [Fig. S2(d)]. At EPs, the eigenvectors are not orthogonal [see Fig. S2(e)], Instead, not

only the eigenvalues, but also the eigenvectors are degenerate [see Fig. S2(f)]. For another non-Hermitian case with $\gamma$=1, $c_2/c_1$=-1.5, it is interesting to find that the two EPs merge and form a single point, and the band structure [Fig. S2(g)] is almost the same as the Hermitian case that features a linear relationship around $k$=0. It should be mentioned that though the similarities, the eigenmode properties of the single point in the non-Hermitian case are different from the single point in Hermitian case [see Figs. S2(h) and S2(i)]. As is clearly displayed, the eigenvectors are degenerate, not orthogonal at the point, which demonstrates an EP property.

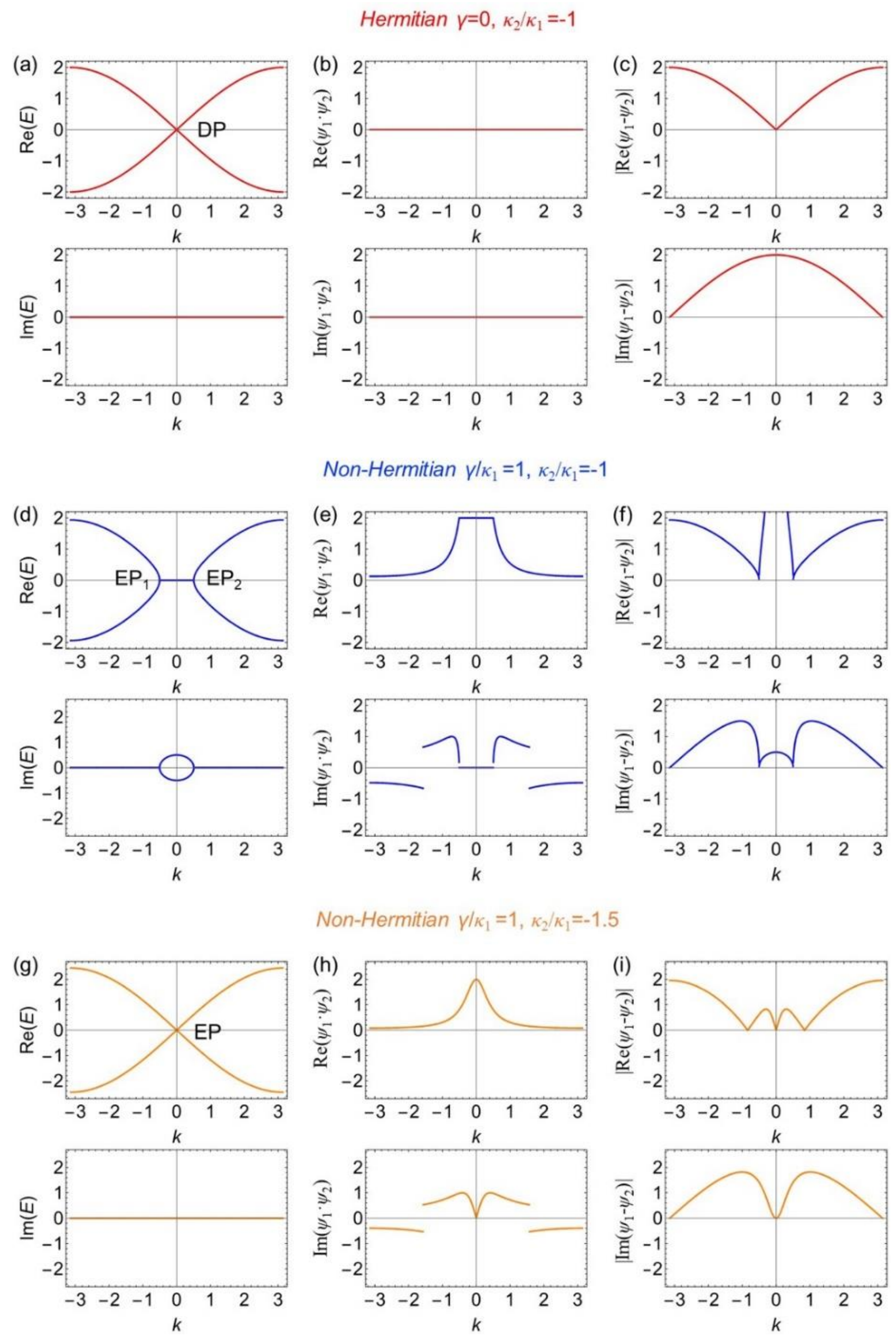

Figure S2 Eigenvalue and eigenmode properties for Hermitian and non-Hermitian systems. (a,d,g) Band structures. (b,e,h) dot product of eigenmodes of two bands ($\psi_1$. $\psi_2$). (c,f,i) $|\psi_1$-$\psi_2|$ in Hermitian and non-Hermitian waveguide lattice.

## II. Analogy between the Hamiltonian of the waveguide lattice and Dirac Hamiltonian

The analogy between Hamiltonian for optical fields in the non-Hermitian waveguide lattice and Dirac Hamiltonian for relativistic quantum particles is valid in the vicinity of $kd$=0. So we retain the expansion terms in $H_{WG}$ (Eq. (S1)) up to the second order of $kd$.

$$H_{WG}^{'}=\begin{pmatrix} i\gamma/2\kappa_1 & 1+\kappa_2/\kappa_1[1-ikd-(kd)^2/2] \\ 1+\kappa_2/\kappa_1[1+ikd-(kd)^2/2] & -i\gamma/2\kappa_1 \end{pmatrix},\quad \text{when } kd \ll 1 \tag{S4}$$

the eigenvalues write

$$\begin{cases} -\dfrac{\Delta^{'}}{2\kappa_1} \\ \dfrac{\Delta^{'}}{2\kappa_1} \end{cases}, \tag{S5}$$

where $\Delta^{'}=\sqrt{4(\kappa_1+\kappa_2)^2-\gamma^2-4\kappa_1\kappa_2(kd)^2}$ , which is

$$\begin{cases} -\sqrt{(\kappa_2/\kappa_1+1)^2-(\gamma/2\kappa_1)^2-\kappa_2/\kappa_1(kd)^2} \\ \sqrt{(\kappa_2/\kappa_1+1)^2-(\gamma/2\kappa_1)^2-\kappa_2/\kappa_1(kd)^2} \end{cases}, \tag{S6}$$

here, the $(kd)^4$ term has been ignored.

The Dirac Hamiltonian was derived by linearizing the famous energy-momentum relation $E^2=m_0^2c^4+p^2c^2$, which is

$$H_D=\begin{pmatrix} mc^2 & cp \\ cp & -mc^2 \end{pmatrix}. \tag{S7}$$

Here, the Dirac-like Hamiltonian form for the waveguides systems $H_a$ can be derived from the eigenvalue Eq. (S6) in a similar manner

$$H_a=\begin{pmatrix}\sqrt{(\kappa_2/\kappa_1+1)^2-(\gamma/2\kappa_1)^2} & \sqrt{-\kappa_2/\kappa_1}kd\\ \sqrt{-\kappa_2/\kappa_1}kd & -\sqrt{(\kappa_2/\kappa_1+1)^2-(\gamma/2\kappa_1)^2}\end{pmatrix}, \tag{S8}$$

Eq. (S8) is valid in the vicinity of $kd=0$. Here, by setting $(-\kappa_2/\kappa_1)^{1/2}\equiv c_a$, $kd\equiv p_a$, and $[(\kappa_2/\kappa_1+1)^2-(\boldsymbol{\gamma}/2\kappa_1)^2]^{1/2}/(-\kappa_2/\kappa_1)\equiv m_a$, Eq. (S8) can be rewritten as

$$H_a=\begin{pmatrix} m_a c_a{}^2 & c_a p_a\\ c_a p_a & -m_a c_a{}^2\end{pmatrix}. \tag{S9}$$

Note that the Hamiltonian of the waveguide lattice [Eq. (S9)] and Dirac Hamiltonian [Eq. (S7)] have the same mathematical form. Thus, the dynamics of relativistic particles governed by Dirac Equation can be mapped to the transfer of a wave packet in a photonic waveguide lattice. The bases of the matrix operator $H_a$ are different from those of $H_{WG}{}'$ in Eq. (S4). However, they produce the same eigenvalues as Eq. (S6). Actually, $H_a$ can be obtained through proper unitary transformation of $H_{WG}$'.

$$H_a=\psi_a\psi_{WG}{}^{-1}H_{WG}{}'\psi_{WG}\psi_a{}^{-1}, \tag{S10}$$

where $\psi_a=(\psi_{a1},\psi_{a2})$, $\psi_{WG}=(\psi_{WG1},\psi_{WG2})$ are eigenstates of $H_a$, $H_{WG}{}'$, respectively.

$$\begin{cases}\left(\dfrac{(i\gamma-\Delta')}{2[\kappa_1+\kappa_2[1+ikd-(kd)^2/2]},1\right)\\ \left(\dfrac{(i\gamma+\Delta')}{2[\kappa_1+\kappa_2[1+ikd-(kd)^2/2]},1\right)\end{cases}, \tag{S11a}$$

$$\begin{cases}\left(\dfrac{(\sqrt{4(\kappa_1+\kappa_2)^2-\gamma^2}-\Delta')}{2\sqrt{-\kappa_1\kappa_2}kd},1\right)\\ \left(\dfrac{(\sqrt{4(\kappa_1+\kappa_2)^2-\gamma^2}+\Delta')}{2\sqrt{-\kappa_1\kappa_2}kd},1\right)\end{cases}, \tag{S11b}$$

**III. Band structures under different excitation conditions**

The similarities between the paraxial wave equation and the Schrodinger equation inspire us to solve Eq. (1) in the main text in a quantum way. Under the tight-binding approximation, the matrix form of the Hamiltonian of the waveguide lattice (81 waveguides) consists only diagonal and off-diagonal elements

$$H_{81\times 81}=\begin{pmatrix} \beta_0-i\gamma/2 & \kappa_1 & & & & & \\ \kappa_1 & \beta_0+i\gamma/2 & \kappa_2 & & & & \\ & \kappa_2 & \ddots & \ddots & & & \\ & & \ddots & \ddots & \kappa_1 & & \\ & & & \kappa_1 & \beta_0+i\gamma/2 & \kappa_2 & \\ & & & & \kappa_2 & \beta_0-i\gamma/2 \end{pmatrix}_{81\times 81}. \tag{S12}$$

The eigenvalues of Eq. (S12) is denoted by $\beta_i$ and the corresponding eigenvectors are $\psi_{Ei}$ = ($\psi_{i1}$, $\psi_{i2}$, ..., $\psi_{i81}$)$^T$, where $i$ = 1, 2, ..., 81 is the mode number. We consider our initial input wave packet $\psi_{in}$ =(0, …, exp(-$i$5$\varphi_0$), …, exp(-$i\varphi_0$), exp(-$i\varphi_0$/2), 1, exp($i\varphi_0$/2), exp($i\varphi_0$), …, exp($i$5$\varphi_0$), …, 0)$_{81\times 81}$ for the central 21 waveguide excitation that controls distribution of the momentum centered at $\varphi_0$, and project $\psi_{in}$ into all eigenstates $\psi_E$=($\psi_{E1}$, $\psi_{E2}$, ..., $\psi_{Ei}$, …, $\psi_{E81}$). A series of occupation weight for $\psi_{in}$ onto each eigenstate $\psi_i$ can be obtained

$$p=\psi_{in}\cdot\psi_E=(p_1,p_2,\cdots,p_i,\cdots,p_{80},p_{81}). \tag{S13}$$

By performing Fourier transform of the field distribution with the eigenvectors in their weight factors, we obtain an expression of excitation intensity on $k$ components in the band structure as

$$\psi_{Ei}(k)=p_i\sum_n \psi_{Ei}(n)\exp(-ink), \tag{S14}$$

where $n$ is the lattice number. Thus, we can draw the band structures of the waveguide lattice under different excitation conditions.

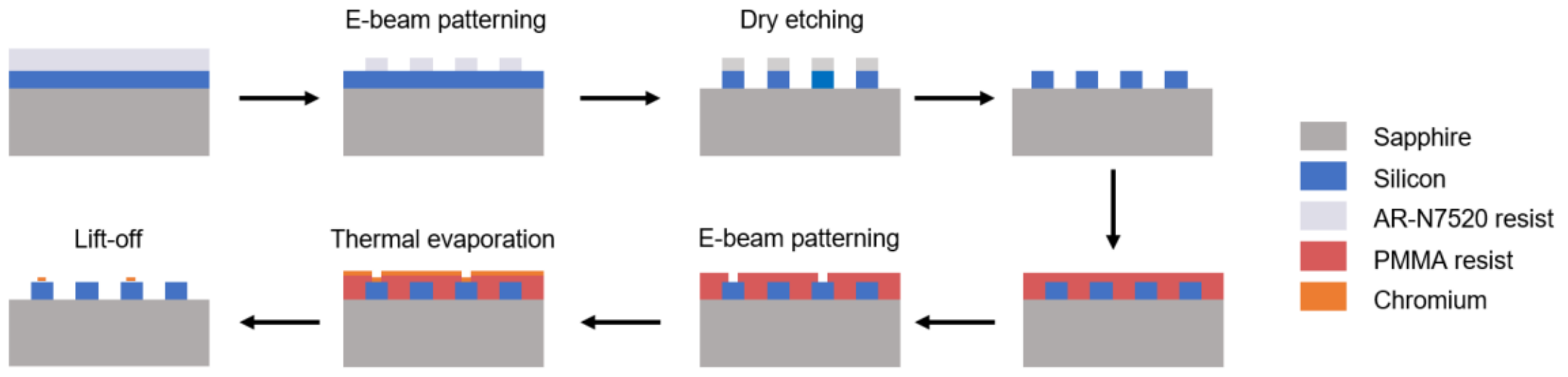

Figure S3 Fabrication flow of the photonic silicon waveguide array deposited with Cr.

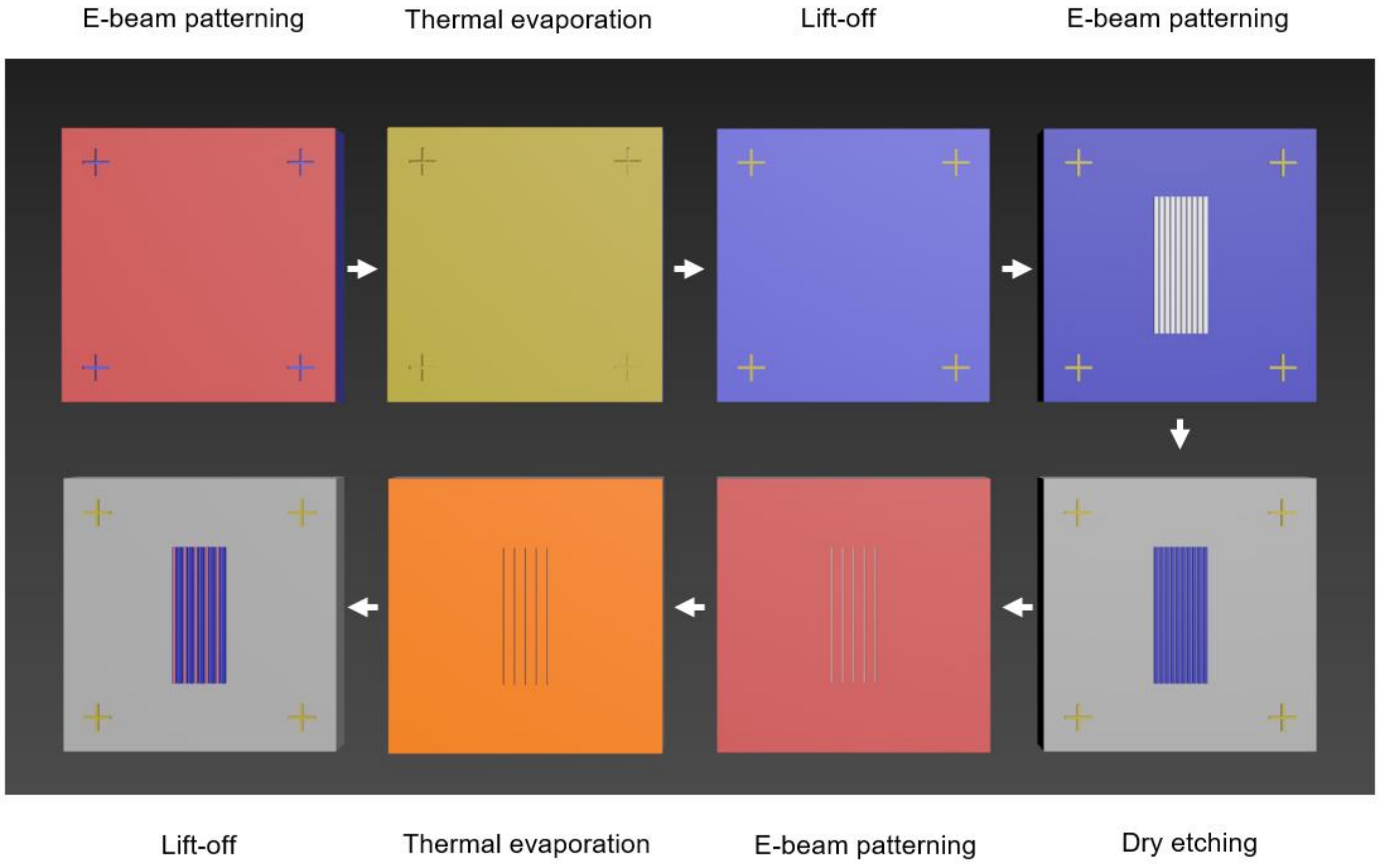

Figure S4 Sample schematics of each step.

## IV. Input branched waveguides design

To ensure a well-defined $k$ momentum, a broad beam that covers several waveguides should be launched. In experiments, we input light into the grating coupler to couple the light into a single waveguide mode, then a branched waveguide is used to splitter the light into several waveguides that are directly connected to the waveguide array with specific phase relationship. The specific value of the momentum $k$ is controlled by optical path difference of different branched waveguides, which results in phase difference. According to the simulations, three waveguides are enough to distinguish the dynamics of different particles. In experiments, we designed trident branched waveguides that consists of three branch waveguides, one is straight and the other two are half-period sinusoidal curves with different amplitude $A_1$ ($A_2$) and period $2L_1$($2L_2$), as the schematics shown in Fig. S5(c). $A_1$ ($A_2$) and period $2L_1$($2L_2$) should satisfy the following conditions

$$A_1=d_1/2, \tag{S15a}$$

$$A_2=d_2/2, \tag{S15b}$$

$$\beta_1 s_1(A_1, L_1)- \beta_2 s_2(A_2, L_2)-\beta_0(L_1-L_2)=\varphi_0, \tag{S15c}$$

where $d_1$ and $d_2$ are the spacing between adjacent waveguides. $\beta_1$ and $\beta_2$ are propagation constants of the two curved waveguides, which can be reasonably treated as the same as the straight waveguide $\beta_0$, i.e. $\beta_1$=$\beta_2$=$\beta_0$, as the waveguides are weakly curved. $s_1(A_1, L_1)$ and $s_2(A_2, L_2)$ are the length of curved waveguides, which is calculated by a curvilinear integral $\boldsymbol{s_j(A_j, L_j) = \int_0^{L_j} \sqrt{1+\left(\frac{A_j\pi}{L_j}\right)^2 cos^2(\frac{\pi}{L_j}x)}dx}$, where $j$=1,2. $\varphi_0$ is the phase difference of input optical fields across one structure period. In our design, $d_1$=593 nm, and $d_2$=880 nm, and $\beta_0$=2.16$k_0$=8.76 μm$^{-1}$. Based on these parameters, $A_1$=296.5 nm, $A_2$=440.0 nm. By taking $L_1$=20 μm, $\varphi_0$ can be uniquely determined by choosing different $L_2$. $\varphi_0$-$L_2$ relationship was simulated through a commercial finite-element analysis solver (COMSOL MULTIPHYSICS 5.3). Figure S5(d) shows $\varphi_0$ as a function of $L_2$. Two $L_2$ are selected, i.e., $L_2$=15.1 μm for $\varphi_0$=0 and $L_2$=16.0 μm for $\varphi_0$=0.6. Figures S5(e,f) show the simulated phase arg($E_x$) as the light propagation through the trident branched waveguide with $L_2$=15.1 μm (e) and $L_2$=16.0 μm (f). The specific $\varphi_0$ can be obtained at the end of the branched waveguide that is directly connected to the corresponding waveguide arrays. The experimental samples are fabricated according to the simulated parameters, and the SEM pictures are shown in Figs. S5(a,b).

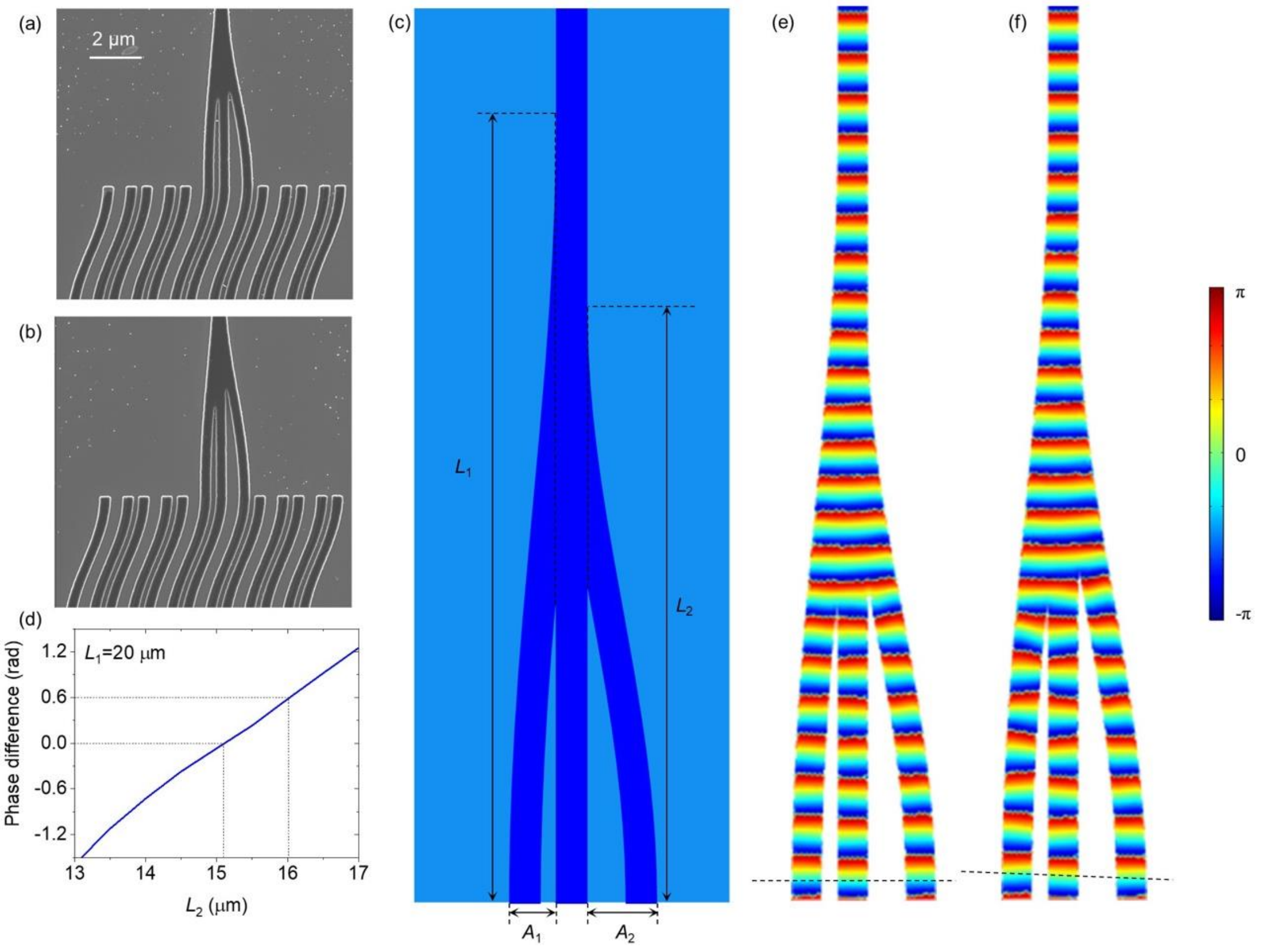

Figure S5 Input branched waveguide design. (a,b) SEM pictures of the input trident branched waveguide. (c) Schematics of the designed input port with three branch waveguides. (d) $\varphi_0$ as a function of $L_2$. (e,f) Simulation results of the field evolution in the designed input port.

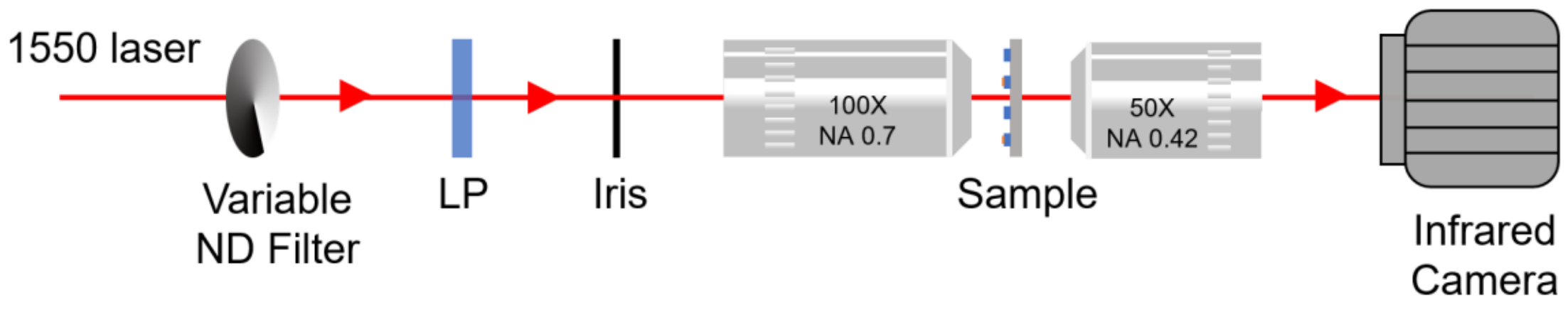

Figure S6 Experimental setup to characterize the dynamics of relativistic particles.

## V. Dynamics of non-Hermitian massless particles

Additionally, as we mentioned before, the massless particles can also exist in non-Hermitian systems (see the white dashed line in Fig. 1(c)). As an example, we demonstrate the non-Hermitian massless particle with $\gamma/\kappa_1$=1, $\kappa_2/\kappa_1$=-0.5. Figures S7(a,c) show the excited band structures, where the gray dashed lines represent the band structures and yellow dashed lines mark the excitation positions. The corresponding optical field evolutions are shown in Figs. S7(b,d). Note that though these band structures are similar (see gray dashed line in Figs. S7(a,c)) with featured linear dispersion, the optical field evolutions are completely different. The beam exhibits classical conical diffractions in Hermitian systems while demonstrates anomalous constant intensity conical diffraction in non-Hermitian systems [44,45] for $\varphi_0$ centered at $p_a$=0 (see theoretical, simulation and experimental results in Figs. S7(d-f)). The anomalous phenomena around the $p_a$=0 arise from the non-Hermiticity of EPs. To be more specific, the Petermann factor, an important observable that measures the degree of non-Hermiticity of the system, diverges to infinity at the EP while it equals to unity at the DP [46,47]. These results indicate that the massless particles in non-Hermitian systems evolve quite differently from that in Hermitian systems.

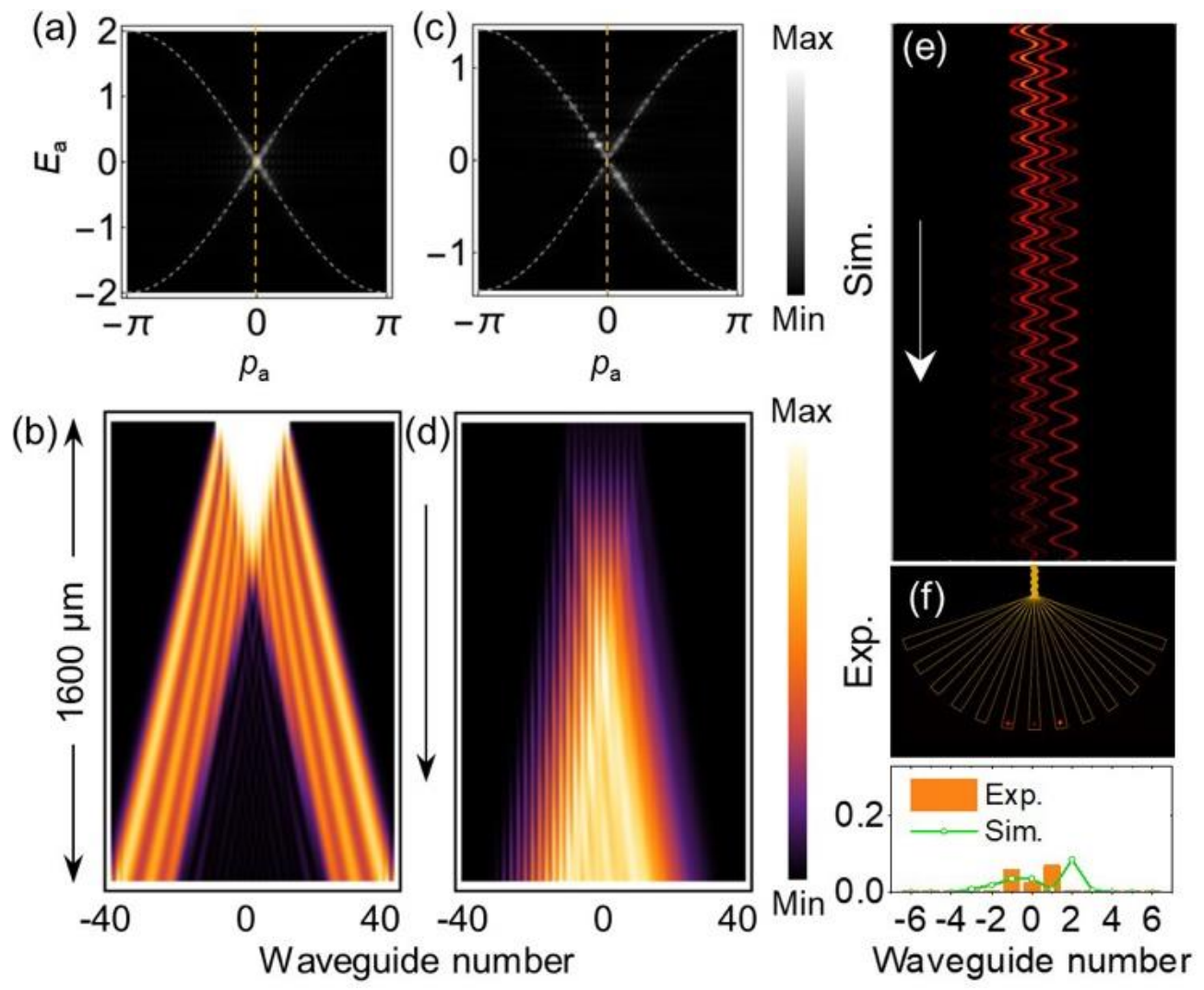

Figure S7 Demonstrations of non-Hermitian massless particles. Excited band structures for massless particles in Hermitian (a) and non-Hermitian (c) systems, where the gray dashed lines guide the band structures, the yellow dashed lines mark the excitation position. (b,d) The corresponding optical field

evolution patterns. (e) Simulated optical field dynamics for massless particles in non-Hermitian systems. (f) Corresponding experimentally detected output intensities (top) and normalized intensity profiles (bottom).